\documentclass[prl,twocolumn,showpacs,preprintnumbers,amsmath,amssymb]{revtex4}
\usepackage{amsmath}
\usepackage{graphics}
\usepackage{epsf}
\usepackage{dcolumn}
\usepackage{bm}
\begin{document}

\title{Enhanced spin density wave in LaOFeSb}

\author{Chang-Youn Moon, Se Young Park, and Hyoung Joon Choi}
\email[Email:\ ]{h.j.choi@yonsei.ac.kr}
\affiliation{Department of Physics and IPAP, Yonsei University, Seoul 120-749, Korea}

\date{\today}

\begin{abstract}
We predict atomic, electronic, and magnetic structures of a hypothetical compound LaOFeSb by first-principles 
density-functional
calculations. It is shown that LaOFeSb prefers a stripe-type antiferromagnetic 
phase (i.e., spin density wave (SDW) phase) to the non-magnetic (NM) phase, with a larger Fe spin moment 
and greater SDW-NM energy difference than those of LaOFeAs. The SDW phase is found to 
favor the orthorhombic structure while the tetragonal structure is more stable in the NM phase.
In the NM-phase LaOFeSb, the electronic bandwidth near the Fermi energy is reduced compared with LaOFeAs, 
indicating smaller 
orbital overlap between Fe $d$ states and subsequently enhanced intra-atomic exchange coupling.
The calculated Fermi surface in the NM phase consists of three hole and two electron sheets, and shows
increased nesting between two hole and two electron sheets compared with LaOFeAs. 
Monotonous changes found in our calculated material properties of LaOFePn (Pn=P, As, and Sb),
along with reported superconducting properties of doped LaOFeP and LaOFeAs, suggest
that doped LaOFeSb may have a higher superconducting transition temperature.
\end{abstract}


\maketitle
Since the discovery of a new class of superconductor LaOFeP with a superconducting transition temperature
$T_{c} \sim $ 4 K \cite{Kamihara2006} and subsequent dramatic increase to 26 K with LaO$_{1-x}$F$_x$FeAs 
compound \cite{Kamihara2008}, and to 43 K with external pressure \cite{Takahashi}, a huge amount of studies have followed to 
understand the underlying 
mechanism of the superconductivity in these iron-based layered oxypnictide compounds and to raise 
$T_c$ further. Often being compared with cuprate
superconductors, this family of materials is expected to give a clue to the superconductivity
in the cuprates, with similarities such as two-dimensionality of structural and electronic properties,
exhibition of the antiferromagnetic (AFM) ground state of the undoped parent materials, subsequent
loss of the magnetism accompanied by the emergence of the superconductivity with doping, etc.
As the magnetism is believed to be closely related to the superconductivity of these materials,
many studies have been focused on the magnetic property of the materials, 
and a number of theoretical studies have predicted magnetic instability in pure LaOFeAs 
\cite{Giovannetti,Singh,Haule,Xu,Mazin}. 
It is known that there is a structural phase transition from tetragonal to orthorhombic phase 
near 150 K \cite{Cruz,Klauss}, which is associated with anomalies in resistivity, dc magnetic susceptibility,
and specific heat \cite{Kamihara2008,Dong}. At temperatures below $\sim$ 130 K, an AFM
order of spin density wave (SDW) type was identified with magnetic moments of about 0.3 $\mu_B$ 
\cite{Cruz,Klauss}.
The existence of SDW state in these materials have been confirmed in many other experimental \cite{Nakai,Lorenz,Dong} 
and theoretical \cite{Dong,Cvetkovic,Korshunov,Yin} works, and can be understood by a peculiar 
Fermi surface topology, i.e., the Fermi surface nesting.

Since the conducting FeAs layer 
is conceived to be the relevant part of magnetism and superconductivity, there have been large efforts to tailor
the structure by modifying the insulating layer, which is LaO layer in the case of LaOFeAs, to raise $T_c$.
The constraint considered is 
that the insulating layer should provide one electron per one formula unit of FeAs and substantial doping
should be introduced by substitution or non-stoichiometry. By substituting La with other rare-earth (RE) 
elements, series of RE(O$_{1-x}$F$_x$)FeAs 
are reported to have $T_c$ of 55 K for RE=Sm \cite{Ren2053}, 41 K for RE=Ce \cite{Chen3790}, 52 K for RE=Pr 
\cite{Ren4283}, and 50 K for RE=Nd \cite{4234}. Besides the fluorine doping, 
$T_c$ of 53.5 K is reported with GdOFeAs with oxygen deficiency \cite{Yang}. 
More recently, BaFe$_2$As$_2$, which contains the same FeAs layer as in the REOFeAs
compounds but Ba layer instead of REO layer, has been found to have similar properties of magnetism 
and superconductivity, with the SDW transition at 140 K in undoped samples \cite{0805.4021}, and superconducting 
transition at 38 K in K-doped ones \cite{0805.4630}. SrFe$_2$As$_2$ is another similar discovery, with the SDW
transition at 205 K \cite{0806.1043,0806.1209} and superconducting $T_c$ of about 38 K \cite{0806.1209} 
for K-doped samples. In these many materials, variations
made on the material so far have been restricted to the insulating layer except for FeSe cases, 
preserving the FeAs layer itself since the first breakthrough of substituting P of LaOFeP with As \cite{Kamihara2008}.
A straightforward idea to modify directly the FeAs layer is substituting As with Sb, inspired by 
the great success of As replacing P.

In this work, we study the structural, electronic, and magnetic properties of a hypothetical iron-based 
oxypnictide compound LaOFeSb through first-principles density-functional pseudopotential calculations. 
We find that the SDW-type AFM spin state is the ground state for this compound, similar to existing other 
oxypnictides LaOFeP and 
LaOFeAs. The crystal structure couples with the magnetic configuration so that the NM phase favors the 
tetragonal structure whereas the lower-symmetry orthorhombic structure is
favored by the SDW state. As the distribution of Fe $d$ bands near the Fermi level ($E_f$) is different
from that of LaOFeAs, the Fermi surface topology
is also different. The Fermi surface consists of three hole pockets around the $\Gamma$ point and 
two electron pockets around
the M points, with enhanced nesting between the hole and electron surfaces. 
We find that LaOFeSb in the SDW phase has a local Fe spin moment larger than that of LaOFeAs, making a trend
of monotonous increase from phosphide to antimonide, along with the relative stability of the magnetic
ground state with respect to the NM solution. 
Our calculation suggests LaOFeSb, which is not yet reported to be synthesized, could be a candidate
for higher $T_c$.

\begin{figure}
\epsfxsize 3.5in
\centerline{\epsffile{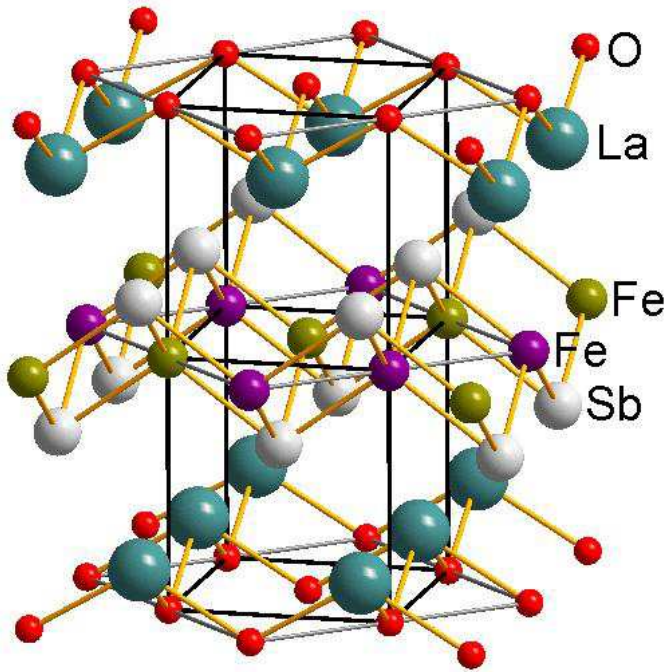}}
\caption{(color online). Atomic structure of LaOFeSb. In this structure, LaO and FeSb layers alternate
along the c direction. Black lines represent the tetragonal unit cell of the NM phase, and $\sqrt{2} 
\times \sqrt{2}$ supercell is also indicated in gray lines. The Fe spin configuration in the SDW phase
is shown with different colors of Fe atoms.
}
\label{fig1}
\end{figure}

Our first-principles calculations are based on the density-functional theory (DFT) within
the generalized gradient approximation (GGA) for the exchange-correlation energy functional
\cite{PBE} and the {\it ab-initio} norm-conserving pseudopotentials as implemented in SIESTA 
code \cite{SIESTA}. Semicore pseudopotentials are generated using electronic configurations
$5s^25p^25d^06s^0$ with $5s$ and $5p$ electrons as valence electrons, and $3s^23p^63d^64s^0$
with $3s$, $3p$, and $3d$ electrons as valence electrons, for La and Fe, respectively.
Electronic wave functions are expanded with localized pseudoatomic orbitals (double zeta polarization
basis set), with the cutoff energy for real space mesh of 400 Ry. Brillouin zone integration is
performed by Monkhorst-Pack scheme \cite{Monkhorst} with 12 $\times$ 12 $\times$ 6 k-point grid.
The unit-cell parameters are optimized by the total energy minimization and used for the electronic 
structure calculations.
Atomic forces are relaxed below 0.04 eV/\AA. When our calculational method is applied to LaOZnPn 
(Pn=P, As, and Sb), as a test, the obtained structural parameters are in good agreement with experimentally
known values. 

Oxypnictide compounds have layered structure with alternating LaO and FePn layers along
the c-axis, as shown in Fig. 1. In the FePn layer Fe atoms form a two-dimensional square lattice and 
are four-fold coordinated with Pn atoms to form a network of tetrahedrons, and the LaO layer 
is in structural analogy with FePn layer with Fe and Pn atoms replaced by O and La atoms, respectively. 
LaOFeAs is known to have two structural phases of different symmetries with a transition temperature 
of $\sim$ 150 K \cite{Cruz,Klauss} as mentioned earlier. The higher temperature phase has tetragonal $P4/nmm$
symmetry and the lower temperature one has monoclinic $P112/n$, or equivalently, orthorhombic $Cmma$ 
\cite{Yildirim}. 

To optimize the structure of LaOFeSb as well as LaOFeP and LaOFeAs, we choose a 
$\sqrt{2} \times \sqrt{2} \times 2$ supercell 
of the original tetragonal (1 $\times$ 1 $\times$ 1) primitive unit cell for the starting 
structure and fully relax the structure including the cell parameters. We choose the doubled
unit cell along the c-axis to examine
the experimentally observed spin configuration in LaOFeAs that the spin direction is opposite between 
two adjacent Fe layers along the c-axis \cite{Cruz}. With this choice of the supercell, we perform both 
the NM calculation and spin-polarized one with the well-known SDW spin configuration as represented in 
Fig. 1.

\begin{table}
\caption{Calculated structural parameters, DOS at $E_f$ ($N(E_f)$), and Fe magnetic moment (m) of 
LaOFePn, for Pn=P, As, and Sb.
Optimized crystal structure for NM phase is tetragonal with 8 atoms in a unit cell
and that for the SDW phase is orthorhombic with 16 atoms in a unit cell. Spin ordering
in the SDW phase doubles the unit cell size along the c axis with 32 atoms in a unit cell.
The lattice parameters $a_o$ and $b_o$ in the orthorhombic structure are slightly greater
than $\sqrt{2}$ times of $a$ in the tetragonal structure. In the SDW phase, the Fe atoms
of the same spin are located along the $b_o$ axis. Iron (Oxygen) atoms are located at
$z = 0.5~(z = 0)$ along the c-axis.
}
\begin{center}
\begin{tabular}{ccccccc}
\hline\hline
\multicolumn{7}{c}{NM (Tetragonal, $P4/nmm$) }  \\
 Pn  & ~~$a$ (\AA)~~   & ~~$b$ (\AA)~~ &  ~~$c$ (\AA)~~   &  ~~$z$(La)~~  & ~~$z$(Pn)~~ & $N(E_f)$\\
\hline
 P     &  3.972 & 3.972 &  8.590   &  0.153 &  0.624  & 4.5 \\
 As     &  3.999 & 3.999 &  8.706  &  0.145 &  0.640  & 1.7 \\
 Sb     &  4.106 & 4.106 &  9.311  &  0.130 &  0.659  & 2.9 \\
\hline
\multicolumn{7}{c}{SDW (Orthorhombic, $Cmma$) }  \\
 Pn  & $a_o$ (\AA)   & $b_o$ (\AA) &  $c$ (\AA)   &  $z$(La)  &  $z$(Pn) &  m($\mu_B$)\\
\hline
 P      &  5.741 & 5.672 &  8.641  &  0.150 &  0.634  & 2.42  \\
 As     &  5.780 & 5.693 &  8.875  &  0.139 &  0.654  & 2.83  \\
 Sb     &  5.955 & 5.844 &  9.542  &  0.124 &  0.673  & 3.13  \\
\hline\hline
\end{tabular}
\end{center}
\label{table I}
\end{table}
The optimized cell parameters and atomic coordinates of LaOFeSb are listed in Table I along 
with those for LaOFeP and LaOFeAs for comparison. For all of the compounds, our NM calculations 
result in tetragonal structures with $a=b$, and the lattice parameters of LaOFeSb are slightly
larger than those of LaOFeAs as expected. The NM-phase tetragonal structure is not stable with respect
to AFM spin orderings. Our spin-polarized calculation for the SDW phase leads to
an orthorhombic unit cell with $a \neq b$, and again LaOFeSb is found to have larger lattice parameters
than LaOFeAs. Our result for LaOFeAs is consistent with the previous
calculation \cite{Yildirim} which shows that, in the SDW spin configuration, parallel spins tend to get 
closer and opposite spins move
apart by structural distortion with slightly increased $\gamma$ angle (between a- and b-axis in 
the original 1 $\times$ 1 unit cell) from 90 degrees to lift the magnetic frustration.
The $\gamma $ angles from our structure optimization are found to be 90.7, 90.9, and 91.1 degrees,
for Pn=P, As, and Sb, respectively. The angle of 90.9 degrees for LaOFeAs is in reasonable agreement 
with the measured value of 90.3 degrees \cite{Cruz}, supporting the validity of our predicted value 
for LaOFeSb.
Moreover, the increasing $\gamma$ angle from P to Sb is also consistent with the fact that
the calculated Fe spin moment increases monotonically from P to Sb, as presented below.

The AFM spin configuration (Fig. 1) in the optimized orthorhombic 
unit cell is more stable than the NM phase in the optimized tetragonal unit cell by 153, 354, and 
706 meV per formula unit for phosphide, arsenide, and antimonide, respectively. 
The local magnetic moment on a Fe atom is also increasing as 2.42, 2.83, and 3.13 $\mu_B$.
Therefore, the magnetism is greatly enhanced in the LaOFeSb compared with LaOFeAs and LaOFeP.
In the meanwhile, no ferromagnetic (FM) solution is found in our calculation. 

Figure 2 shows the band structures calculated in the NM phase for the series of oxypnictides.
Overall features are similar among Pn species. Complicated Fe $d$-derived bands exist
near $E_f$, electron and hole pockets around M and $\Gamma$ points, respectively,
and dispersionless bands between $\Gamma$ and Z points, indicating the two-dimensionality
of the layered structure. However, band widths and 
detailed energy positions of Fe $d$-derived bands are different among pnictides. 

\begin{figure}
\epsfxsize 3.5in
\centerline{\epsffile{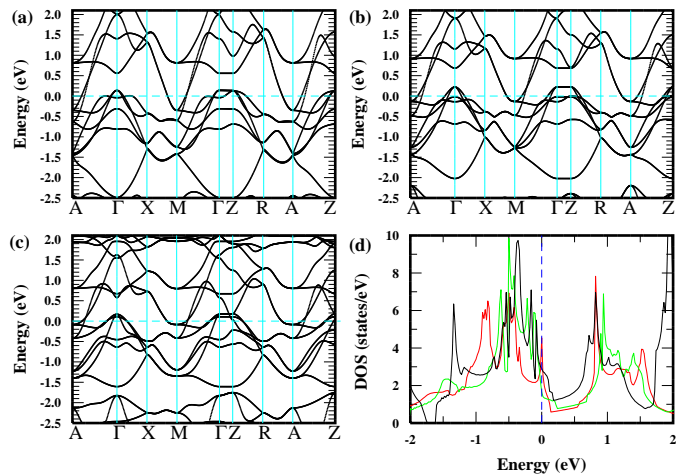}}
\caption{(color online). Calculated band structures of a) LaOFeP, b) LaOFeAs, and c) LaOFeSb for NM
phase. In (d), DOS are shown in red, green, and black colors for LaOFeP, LaOFeAs, and
LaOFeSb, respectively.
}
\label{fig2}
\end{figure}
As shown in Fig. 2, the band dispersion is considerably different among the compounds. 
The P compound has largest dispersion and band width, and then follow the As and Sb 
compounds, as is easily seen, for examples, for the electron bands at M point near $E_f$.
The decreasing dispersion from P to Sb compounds are easily understood by the increasing
Pn atomic size and distance between Fe and neighboring Pn atoms, which results in reduced orbital overlap 
between them and therefore the localization of Fe $d$ states.
The localization of Fe $d$ states close to the atomic orbital limit enhances the intra-atomic
exchange coupling, or, the Hund's rule coupling, so that the Fe local spin moment and the stability
of magnetic solution are expected to increase from P to Sb. Our results presented above are consistent
with this expectation. 

Detailed energy positions of Fe $d$-derived bands are different among pnictides (Fig. 2), and this
affects the number of hole pockets around the $\Gamma$ point.
Energy splittings between two $d_{z^2}$-derived bands (which have finite 
dispersion along the $\Gamma$-Z line), and between $d_{xy}$ and $d_{yz}+d_{zx}$ derived bands (which are 
dispersionless along the $\Gamma$-Z line) decrease from P to Sb. As a result, there are three bands 
crossing the
$E_f$ near the $\Gamma$ point in the case of LaOFeSb while there are only two bands in the cases of LaOFeP
and LaOFeAs. The 3-D hole pocket centered around the Z point, which is present in LaOFeP
and LaOFeAs, is absent in LaOFeSb. This will be discussed in detail later with Fig. 3.

Figure 2(d) shows the density of states (DOS) of LaOFeP, LaOFeAs, and LaOFeSb in the NM phase.
The value of DOS at $E_f$ is 2.9 states/eV per formula unit in LaOFeSb, which is much greater
than 1.7 states/eV in LaOFeAs. This implies that SDW formation and superconductivity may occur
more effectively in LaOFeSb than LaOFeAs, since large DOS at $E_f$ often results in stronger instability
of the Fermi surface. In the case of LaOFeP, DOS at $E_f$ is as large as 4.5 states/eV due to
a very narrow peak right at $E_F$ originating from van Hove singularities which are away from
the $\Gamma$ or M point, so substantial part of DOS may not contribute strongly to the superconductivity 
within the spin-fluctuation scenario. In the case of the SDW phase, the electronic structure 
near $E_f$ is drastically changed from the NM phase. The DOS at $E_f$
is greatly reduced to 0.23 states/eV for the SDW phase due to the SDW gap formation. The SDW phase is still
metallic, however, with a finite DOS value at $E_f$, as the nesting is not perfect as in the case
of LaOFeAs system \cite{Dong}.

\begin{figure}
\epsfxsize 3.5in
\centerline{\epsffile{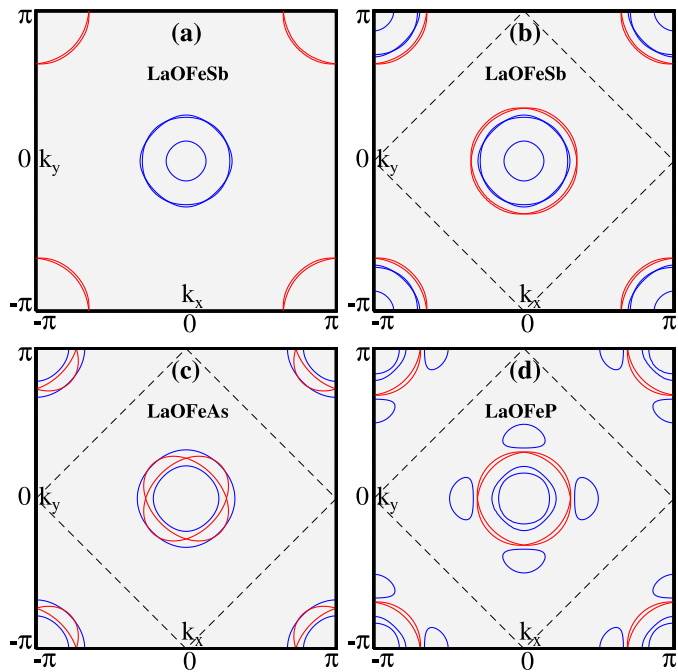}}
\caption{(color online). Calculated Fermi surfaces in the NM phase: (a) LaOFeSb using the primitive
unit cell, and (b) LaOFeSb, (c) LaOFeAs, and (d) LaOFeP using the $\sqrt{2} \times \sqrt{2}$ supercell.
Hole pockets are represented in blue, and electron pockets are in red. In (b)-(d), the dashed lines
represent the Brillouin zone of the $\sqrt{2} \times \sqrt{2}$ supercell.
}
\label{fig3}
\end{figure}

The Fermi surface of LaOFeSb is shown in Fig. 3 on the $k_z=0$ plane. The Fermi surfaces of LaOFeAs 
and LaOFeP are also shown for comparison. Similarly to the LaOFeAs and LaOFeP cases, there are hole 
pockets around the $\Gamma$ point and electron pockets near the M point in LaOFeSb. In this case, however, 
there is additional
small hole Fermi surface centered at $\Gamma$ which is absent in LaOFeAs and LaOFeP, as a result of 
different Fe-$d$ band distributions near $E_f$ (see Fig. 2). Therefore, LaOFeSb has three hole pockets near the 
$\Gamma$ point and two electron pockets around the M point. All Fermi surfaces are cylindrical. 
The small hole surface at $\Gamma$ has different orbital character from the 3-D hole pocket at
Z in LaOFeAs and LaOFeP. The nesting between electron and hole Fermi surfaces defined
by the ($\pi$,$\pi$) vector in the reciprocal space is also manifest in this case. In order to
make comparison straightforward, Figs. 3(b)-(d) show the original Fermi surfaces and those
shifted by the ($\pm \pi$,$\pm \pi$) vectors. As shown in Fig. 3(b), the shapes of Fermi surfaces
in LaOFeSb are quite close to circles compared with those in LaOFeAs and LaOFeP, and the radii
of circular surfaces match closely between the hole and electron surfaces. Furthermore, the electron 
surfaces coincide with each other almost perfectly, and so do the two hole surfaces at $\Gamma$. 
These features which indicate an enhanced nesting between the surfaces, 
together with the enhanced DOS at $E_f$, suggest 
that this material may have stronger pairing mechanism within the spin fluctuation scenario 
which can, in turn, result in superconductivity with higher $T_c$.

In summary, we have investigated the basic physical properties of a hypothetical iron-based oxypnictide 
compound LaOFeSb by first-principles calculations.
It is predicted that the compound favors to be in the tetragonal crystal structure with $a=b$
in the NM state while orthorhombic structure with $a \neq b$ is the lowest energy structure
in the stripe-type AFM-SDW phase. The SDW phase becomes more stable, having the 
local spin moments on Fe atoms and the total energy difference between the SDW
and NM phase larger than those in LaOFeAs.
In the NM phase, the band dispersion around $E_f$ is substantially reduced, increasing DOS
at $E_f$ greatly. The calculated Fermi surface in the NM-phase LaOFeSb exhibits
enhanced nesting feature between the hole and electron surfaces with the nesting vector 
($\pi$,$\pi$) for the SDW formation.

\begin{acknowledgments}
This work was supported by the KRF (KRF-2007-314-C00075) and 
by the KOSEF Grant No. R01-2007-000-20922-0. Computational resources 
have been provided by KISTI Supercomputing Center (KSC-2008-S02-0004). 
\end{acknowledgments}


\begin{thebibliography}{29}
\expandafter\ifx\csname natexlab\endcsname\relax\def\natexlab#1{#1}\fi
\expandafter\ifx\csname bibnamefont\endcsname\relax
  \def\bibnamefont#1{#1}\fi
\expandafter\ifx\csname bibfnamefont\endcsname\relax
  \def\bibfnamefont#1{#1}\fi
\expandafter\ifx\csname citenamefont\endcsname\relax
  \def\citenamefont#1{#1}\fi
\expandafter\ifx\csname url\endcsname\relax
  \def\url#1{\texttt{#1}}\fi
\expandafter\ifx\csname urlprefix\endcsname\relax\def\urlprefix{URL }\fi
\providecommand{\bibinfo}[2]{#2}
\providecommand{\eprint}[2][]{\url{#2}}

\bibitem[{Kam({\natexlab{a}})}]{Kamihara2006}
\bibinfo{note}{Y. Kamihara {\it et al.}, J. Am. Chem. Soc. {\bf 128}, 10012,
  (2006).}

\bibitem[{Kam({\natexlab{b}})}]{Kamihara2008}
\bibinfo{note}{Y. Kamihara, T. Watanabe, M. Hirano, and H. Hosono, J. Am. Chem.
  Soc. {\bf 130}, 3296, (2008).}

\bibitem[{Tak()}]{Takahashi}
\bibinfo{note}{H. Takahashi {\it et al.}, Nature (London) {\bf 453}, 376
  (2008).}

\bibitem[{Gio()}]{Giovannetti}
\bibinfo{note}{G. Giovannetti, S. Kumar, and J. van den Brink,
  arXiv:0804.0866v2 [Phys. Rev. B (to be published)].}

\bibitem[{Sin()}]{Singh}
\bibinfo{note}{D. J. Singh and M.-H. Du, Phys. Rev. Lett. {\bf 100}, 237003
  (2008).}

\bibitem[{Hau()}]{Haule}
\bibinfo{note}{K. Haule, J. H. Shim, and G. Kotliar, Phys. Rev. Lett. {\bf
  100}, 226402 (2008).}

\bibitem[{Xu()}]{Xu}
\bibinfo{note}{G. Xu {\it et al.}, Europhys. Lett. {\bf 82}, 67002 (2008).}

\bibitem[{Maz()}]{Mazin}
\bibinfo{note}{I. I. Mazin, M. D. Johannes, L. Boeri, K. Koepernik, and D. J.
  Singh, arXiv:0806.1869v2.}

\bibitem[{Cru()}]{Cruz}
\bibinfo{note}{C. de la Cruz {\it et al.}, Nature (London) {\bf 453}, 899
  (2008).}

\bibitem[{Kla()}]{Klauss}
\bibinfo{note}{H. -H. Klauss {\it et al.}, arXiv:0805.0264v1.}

\bibitem[{Don()}]{Dong}
\bibinfo{note}{J. Dong {\it et al.}, Europhys. Lett. {\bf 83}, 27006 (2008).}

\bibitem[{Nak()}]{Nakai}
\bibinfo{note}{Y. Nakai, K. Ishida, Y. Kamihara, M. Hirano, and H. Hosono,
  arXiv:0804.4765v2.}

\bibitem[{Lor()}]{Lorenz}
\bibinfo{note}{B. Lorenz {\it et al.}, Phys. Rev. B {\bf 78}, 012505 (2008).}

\bibitem[{Cve()}]{Cvetkovic}
\bibinfo{note}{V. Cvetkovic and Z. Tesanovic, arXiv:0804.4678v3.}

\bibitem[{Kor()}]{Korshunov}
\bibinfo{note}{M. M. Korshunov and I. Eremin, arXiv:0804.1793v1.}

\bibitem[{Yin()}]{Yin}
\bibinfo{note}{Z. P. Yin {\it et al.}, Phys. Rev. Lett. {\bf 101}, 047001
  (2008).}

\bibitem[{Ren({\natexlab{a}})}]{Ren2053}
\bibinfo{note}{Z. -A. Ren {\it et al.}, Chin. Phys. Lett. {\bf 25}, 2215
  (2008).}

\bibitem[{Che()}]{Chen3790}
\bibinfo{note}{G. F. Chen {\it et al.}, Phys. Rev. Lett. {\bf 100}, 247002
  (2008).}

\bibitem[{Ren({\natexlab{b}})}]{Ren4283}
\bibinfo{note}{Z. -A. Ren {\it et al.}, Mat. Res. Innovat. {\bf 12}, 1 (2008).}

\bibitem[{423()}]{4234}
\bibinfo{note}{Z. -A. Ren {\it et al.}, Europhys. Lett. {\bf 82}, 57002
  (2008).}

\bibitem[{Yan()}]{Yang}
\bibinfo{note}{J. Yang {\it et al.}, Supercond. Sci. Technol. {\bf 21}, 082001
  (2008).}

\bibitem[{080({\natexlab{a}})}]{0805.4021}
\bibinfo{note}{M. Rotter {\it et al.}, Phys. Rev. B {\bf 78}, 020503(R)
  (2008).}

\bibitem[{080({\natexlab{b}})}]{0805.4630}
\bibinfo{note}{M. Rotter, M. Tegel, and D. Johrendt, arXiv:0805.4630v2.}

\bibitem[{080({\natexlab{c}})}]{0806.1043}
\bibinfo{note}{C. Krellner {\it et al.}, arXiv:0806.1043v1.}

\bibitem[{080({\natexlab{d}})}]{0806.1209}
\bibinfo{note}{G. F. Chen {\it et al.}, Chin. Phys. Lett. {\bf 25}, 3403
  (2008).}

\bibitem[{PBE()}]{PBE}
\bibinfo{note}{J. P. Perdew, K. Burke, and M. Ernzerhof, Phys. Rev. Lett. {\bf
  77}, 3865 (1996).}

\bibitem[{SIE()}]{SIESTA}
\bibinfo{note}{D. Sanchez-Portal {\it et al.}, Int. J. Quantum Chem. {\bf 65},
  453 (1997).}

\bibitem[{Mon()}]{Monkhorst}
\bibinfo{note}{H. J. Monkhorst and J. D. Pack, Phys. Rev. B {\bf 13}, 5188
  (1976).}

\bibitem[{Yil()}]{Yildirim}
\bibinfo{note}{T. Yildirim, Phys. Rev. Lett. {\bf 101}, 057010 (2008).}

\end{thebibliography}

\end{document}